\documentclass[traditabstract]{aa}
\usepackage{graphicx}
\usepackage{txfonts}
\usepackage{url}
\usepackage{natbib}
\usepackage{color}

\newcommand{\source}{IGR J17451--3022}
\newcommand{\msun}{{\rm M}_{\sun}}
\newcommand{\rsun}{{\rm R}_{\sun}}
\newcommand{\xmm}{{\textit{XMM-Newton}}}

\newcommand{\integral}{{\textit{INTEGRAL}}}

\newcommand {\blue}[1] {\textcolor{black}{#1}}

\begin{document}

\title{IGR J17451--3022: constraints on the nature of the donor star}

\author{Andrzej A. Zdziarski
 \inst{1}
 \and Janusz Zi{\'o}{\l}kowski
 \inst{1}
 \and Enrico Bozzo
 \inst{2}
 \and Patryk Pjanka
 \inst{3}
 }

\institute{Nicolaus Copernicus Astronomical Center, Polish Academy of Sciences, Bartycka 18, 00-716 Warszawa, Poland
 \and
 ISDC Data Centre for Astrophysics, Chemin d'Ecogia 16, CH-1290 Versoix, Switzerland
 \and
 Department of Astrophysical Sciences, Princeton University, 4 Ivy Lane, Princeton, NJ 08544, USA
 }

\date{Received 23 March 2016 / Accepted 30 August 2016}

\abstract{We constrain the binary parameters of the eclipsing accreting X-ray binary \source\ and the nature of its donor star. The donor mass, its radius, and the system inclination angle are computed based on the system orbital period and eclipse duration recently reported by Bozzo et al.\ (2016). We find that the donor is most likely a main-sequence star with the mass comprised within the range $\sim$(0.5--$0.8)\msun$ and the radius of $\sim\! 0.7\rsun$. Assuming that the accreting compact object in \source\ is a neutron star, the duration of the nearly total rectangular eclipses yields the inclination angle of the system of $71\degr\lesssim i\lesssim 76\degr$, compatible with the presence of dips in this system. We rule out the presence of either brown or white dwarf. However, we find an alternative possibility that the donor star in \source\ could be a partially stripped giant with a very low mass, $\sim\! 0.2\msun$. This case requires a substantial mass loss prior to the formation of the giant-star He core. According to that solution, the radius would be $\sim\! 0.4\rsun$, at $i\sim 80\degr$. We additionally show that the well-known approximate dependence of the giant-star radius exclusively on its core mass breaks down below $\sim\!0.3\msun$.
}

\maketitle

\section{Introduction}
\label{intro}

\source\ is a transient eclipsing and dipping low-mass X-ray binary (LMXB) discovered by \integral\ on 2014 August 22 in the direction of the Galactic Centre \citep{chenevez14}. An \xmm\ observation carried out during this event allowed us \citep{bozzo16} to accurately measure the eclipse duration ($t_{\rm e}\simeq 822\pm 2$\,s) and refine the estimate of the source orbital period at $P=22620.5^{+2.0}_{-1.8}\,{\rm s}$. The eclipses observed from \source\ are virtually rectangular, displaying sharp ingresses and egresses due to the obscuration of the X-ray source by the donor star. The lack of multi-wavelength follow-up observations during the outburst of \source\ did not allow a direct investigation of the nature of its donor star. The non-detection of either X-ray pulsations or type-I X-ray bursts left also the nature of the compact object in this system poorly constrained, though the properties of the source X-ray spectrum provided an indication in favour of a neutron star (NS). If the accretor is an NS, its mass is probably in the range of $1.26\msun< M_1\la 2.0\msun$, where the lower bound corresponds to the Chandrasekhar limit for a white dwarf minus the collapse binding energy and the upper bound is related to the current highest reliable measurements \citep{demorest10,antoniadis13}. Masses as high as $\sim2.4\msun$ cannot, however, be excluded, as this is the maximum theoretical mass that an NS can achieve \citep{lp10}.

In this work, we extend our previous investigations of \source\ by providing constraints on the nature of its donor under the assumption of either an NS or a black hole (BH) accretor. Our constraints in Sect.\ \ref{pars} are derived from the orbital period and eclipse duration of \source, extending previous studies on similar constraints provided, e.g., for SWIFT\,J1749.4--2807 by \citet{ms10} and \citet{altamirano11}. \blue{In Sect.\ \ref{donor}, we perform evolutionary calculations in order to constrain the nature of the donor in this system.}

\section{Constraints on the binary parameters}
\label{pars}

\begin{figure}
\centerline{\includegraphics[width=8cm]{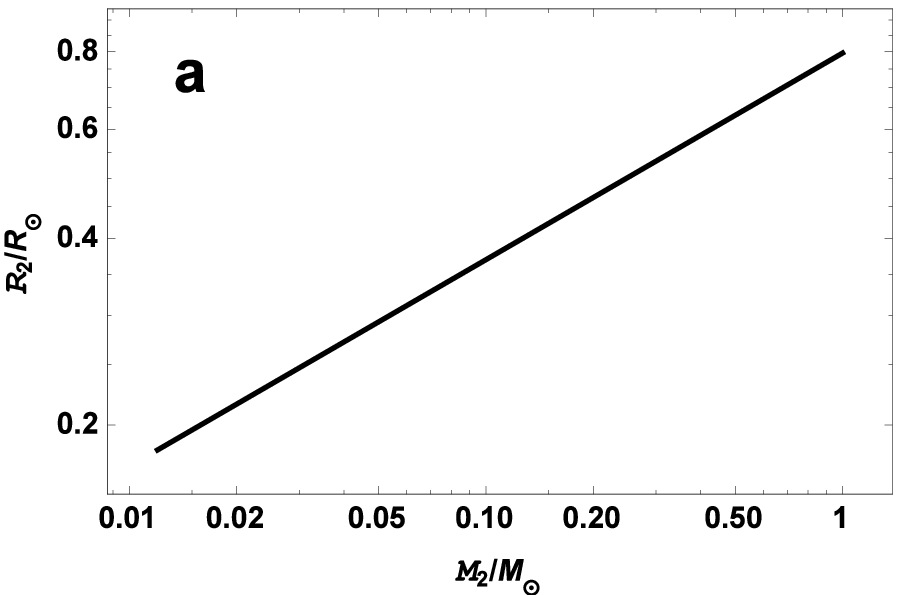}}
\centerline{\includegraphics[width=8cm]{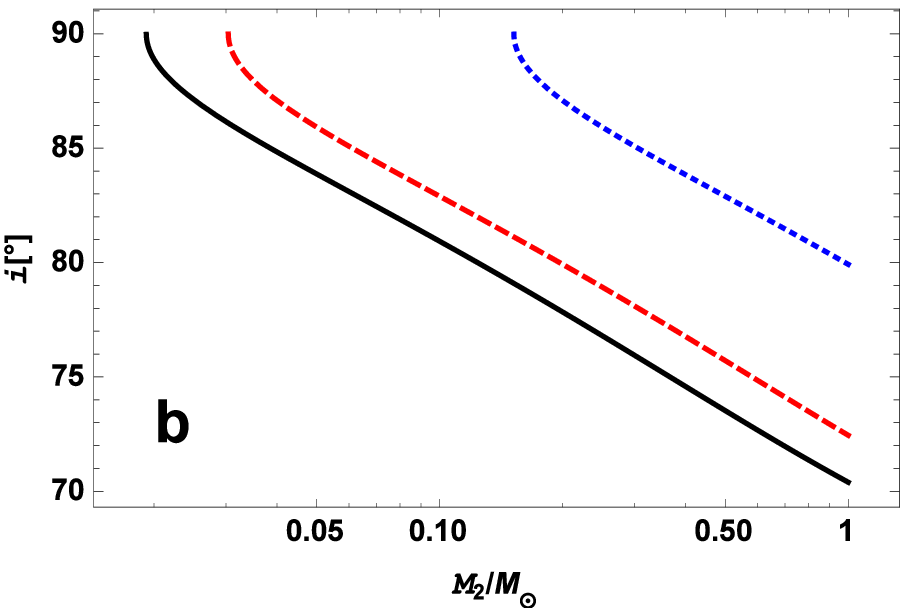}}
\caption{(a) The size of the Roche-lobe radius of the donor as a function of its mass, $M_2$, for the orbital period of $P=22620.5$\,s. As \source\ is an accreting LMXB, $R_2$ is the actual radius of the donor. (b) The relation between the system inclination and the donor mass obtained by using the measured orbital period and eclipse duration. The black solid (red dashed) curve represents the obtained relation in case of an NS accretor with $M_1=1.26\msun$ ($M_1=2\msun$). We also show the relation for a BH accretor of $M_1=10\msun$ (blue dotted curve). } \label{mass}
\end{figure}

As \source\ is an accreting LMXB, the donor star with a mass of $M_2$ should fill its Roche lobe. We can estimate its volume-averaged Roche-lobe radius, $R_2$, using an equation valid in the case $M_2/M_{12}\la 0.6$, where $M_{12}=M_1+M_2$ and $M_1$ is the mass of the compact object. Following \citet{paczynski67}, we have,
\begin{equation}
\frac{R_2}{a}= \frac{2}{3^{4/3}} \left(\frac{M_2}{M_{12}} \right)^{1/3},
\label{roche}
\end{equation}
where $a$ is the semimajor axis of the binary given by the Kepler law,
\begin{equation}
a^3=GM_{12} \frac{P^2}{4\pi^2},
\label{kepler}
\end{equation}
and $G$ is the gravitational constant. The above equations can be combined to obtain a relation between the Roche-lobe radius and the mass of the donor, \begin{equation}
R_2=(2 G M_2)^{1/3}\left(\frac{P}{9\pi}\right)^{2/3}.
\label{r_k}
\end{equation}
\blue{We use hereafter Eqs.\ (\ref{roche}) and (\ref{r_k}) instead of the somewhat more accurate, but more complex, approximation by \citet{eggleton83}. For the range of $M_2/(M_{12})\lesssim 0.4$ that we find possible for \source\ (Sect.\ \ref{donor}), the maximum fractional error of either Eq.\ (\ref{roche}) or (\ref{r_k}) is $\lesssim 1\%$, and quickly decreasing for the decreasing mass ratio.} From Eq.\ (\ref{r_k}), it can be noticed that the average donor star density depends only on $P$ \citep{ffw72}. The $R_2(M_2)$ relationship is plotted in Fig.~\ref{mass}(a). Hereafter, we use the best-fit values of $P$ and $t_{\rm e}$.

By using the measured orbital period and the eclipse duration of \source, we can also relate the system inclination angle, $i$, to $M_2/M_{12}$. At $i=90\degr$, the X-ray source is eclipsed by the equator of the donor star, with the radius $R_2$. At progressively lower inclinations, a correspondingly smaller circle above the equator is eclipsing the compact object. Assuming a circular orbit \citep{chakrabarty93} and neglecting the flattening of the Roche lobe \citep{chanan76}, we have the standard relation,
\begin{equation}
\left(\frac{R_2}{a}\right)^2=\cos^2 i + \sin^2 i\, \sin^2 \frac{\pi\, t_{\rm e}}{P}.
\label{rad_i}
\end{equation}
The above equation can be readily solved for the inclination,
\begin{equation}
i=\frac{1}{2}\arccos{\left[1- 2\frac{1-(R_2/a)^2}{\cos^2 (\pi\, t_{\rm e}/ P)}\right]},\quad \sin \frac{\pi\, t_{\rm e}}{P}\leq\frac{R_2}{a}< 1,
\label{incl}
\end{equation}
which yields $i$ between $90\degr$ and $0\degr$ in the allowed range of $R_2/a$.
From Eq.\ (\ref{incl}), we can obtain a relation between the inclination and the mass ratio by using the dependence of $R_2(M_2)/a(M_{12})$ provided by Eq.\ (\ref{roche}). We show the results in Fig.~\ref{mass}(b) for $M_1=1.26\msun$ and $2\msun$. As a BH accretor in \source\ cannot be completely ruled out (see Sect.~\ref{intro}), we also show the relation $i(M_2)$ for $M_1=10\msun$. We can also determine the minimum allowed donor-to-accretor mass ratio by solving equations (\ref{roche}) and (\ref{rad_i}) at $i=90\degr$, which results in
\begin{equation}
\frac{M_2}{M_{12}}\geq \frac{3^4}{2^3}\sin^3\frac{\pi\, t_{\rm e}}{P}.
\end{equation}
This yields $M_2/M_1\geq 0.0152$.

\section{The nature of the donor}
\label{donor}

\begin{figure}
\centerline{\includegraphics[width=\columnwidth]{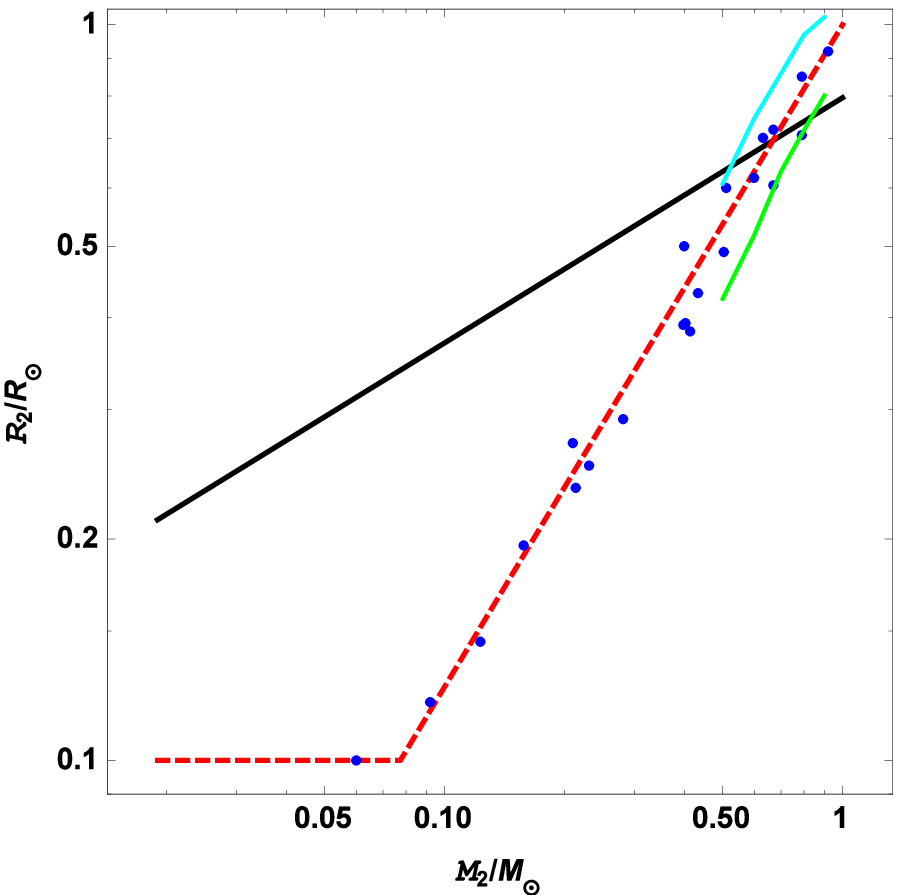}}
\caption{The Roche-lobe radius of the donor star in \source\ as a function of its mass, $M_2$ (black solid line), compared to a selection of the measured masses and radii of stars on the MS (blue points). The dashed red line shows the relation of $R_2=\rsun (M_2/\msun)^{0.9}$ for $M_2\geq 0.1\msun$, which approximates the middle of the MS, and $R_2=0.1\rsun$ for $M_2<0.1\msun$, which approximately describes the case of old brown dwarfs. \blue{The green and cyan curves show the models of stars on the ZAMS and TAMS, respectively. The intersections of these curves with the Roche-lobe radius show that MS stars in the mass range of $\simeq (0.52$--$0.84)\msun$ are acceptable models of the donor star.}
}
\label{radii}
\end{figure}

We can now use the constraints obtained in Sect.~\ref{pars} to investigate different possibilities for the nature of the donor in \source. The most straightforward case is that this star belongs to the main sequence (MS), as it is the case for many other known LMXBs \citep[see, e.g.,][and references therein]{liu07}. In Fig.~\ref{radii} we compare the previously obtained Roche-lobe radius as a function of the donor mass \blue{with the mass-radius relations for MS stars. Given the very small fractional error on $P$ \citep{bozzo16}, the former relationship has virtually no observational error.}

\blue{We made this comparison using theoretical evolutionary models. We use the Warsaw stellar-evolution code (described in \citealt{ziolkowski05}) calibrated to reproduce the Sun at the solar age. This calibration resulted in the chemical composition of the H mass fraction of $X=0.74$, the metallicity of $Z=0.014$, and the mixing length parameter of $\alpha$=1.55. In Fig.~\ref{radii}, we show in the green curve the resulting models for the zero-age MS (ZAMS) and, in the cyan curve, models for stars nearing the end of their MS evolution, at the termination-age MS (TAMS). We find that MS (i.e., central H burning) models with masses in the range $\simeq (0.52$--$0.84)\msun$ could fit the Roche lobe of the mass donor in \source. The lowest and highest mass corresponds to the TAMS and ZAMS, respectively. Models with intermediate masses correspond to intermediate phases of the MS evolution. The corresponding range of the radius is $R_2\simeq (0.64$--$0.75)\rsun$. In addition, the blue points in Fig.~\ref{radii} show a selection of the observationally determined masses and radii of low-mass MS stars from \citet{cox00}, \citet[and references therein]{segransan03} and \citet{pont05}. We see they agree with our theoretical MS bounds.} The inclination corresponding to the obtained allowed mass range is $i\simeq (76$--$71)\degr$ if the donor is an NS, cf.\ Fig.\ \ref{mass}(b). At $M_1=10\msun$, $i\simeq (83$--$81)\degr$.

We then examine alternatives, and first investigate whether the donor in \source\ can be a standard red giant. This could, in principle, be the case if the star had the initial mass $\ga 0.9\msun$; otherwise the evolution within the MS would last longer than the lifetime of the Galaxy \citep[see, e.g.,][]{girardi00}. A low mass donor may be created if a star with a large enough initial mass evolves off the MS, forms a He core, and then loses mass from its H envelope by accretion onto the compact object (see, e.g., \citealt{webbink83}). Such a process leads to formation of so the called partially stripped giants. Donors of this type are present in a number of well-known accreting LMXBs (e.g., GRO 1744--28, GRS 1915+105, and V404 Cyg).

\begin{figure}
\centerline{\includegraphics[width=\columnwidth]{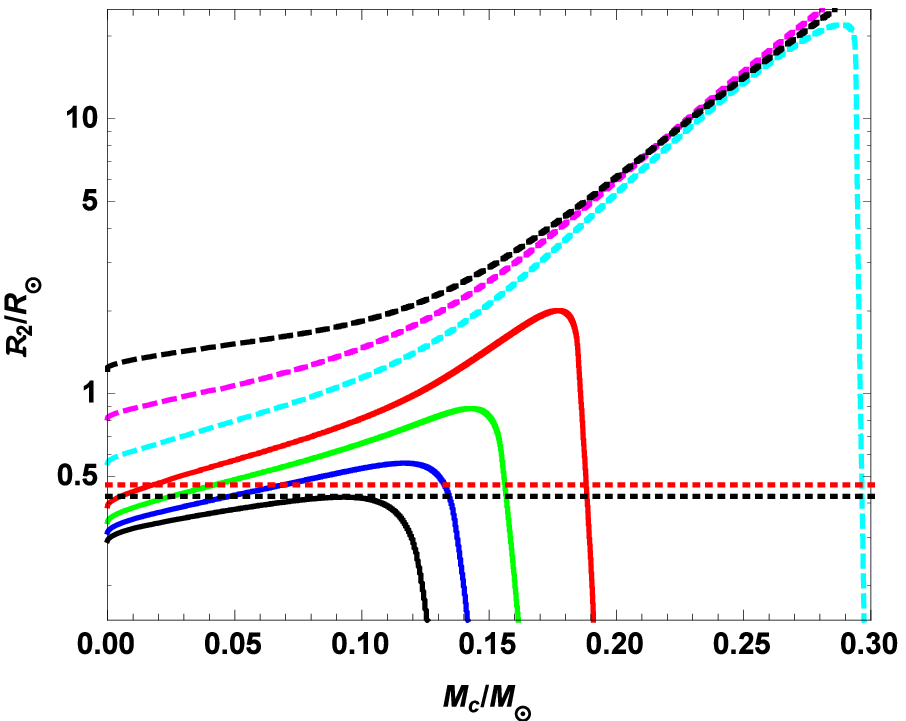}}
\caption{\blue{The evolution of partially stripped giants in the $M_{\rm c}$--$R_2$ diagram. The total star masses are 0.15, 0.16, 0.175, 0.2, 0.3, 0.5 and $0.9\msun$ for the solid black, blue, green and red curves, and dashed cyan, magenta and black curves, respectively. The evolution proceeds (from left to right) at the constant total mass, during which the H-burning shell is moving outwards. This increases the mass of the He core and decreases the mass of the H-rich envelope. The black and red dotted lines show the radius of the Roche-lobe around the donor (and so, to a good approximation, also the radius of this star) for $M_2= 0.15\msun$ and $M_2= 0.2\msun$, respectively. Possible solutions for the donor in \source\ are given by the intersections between the dotted and the solid lines on the left-hand side of the plot.}}
\label{r_mc2}
\end{figure}

The radius of a giant with a total mass $\ga 0.3\msun$ depends primarily on the mass of the core, $M_{\rm c}$ (\citealt{paczynski70}; see also fig.~1 in \citealt{rappaport95}), until its H-rich envelope disappears almost completely. We illustrate it by the three dashed curves in Fig.~\ref{r_mc2}, for the donor masses of 0.3, 0.5 and $0.9\msun$, obtained the Warsaw stellar evolutionary code. As we noted above, the initial mass of the donor has to be $M_2\gtrsim 0.9\msun$. The mass of its core is constrained by the Sch{\"o}nberg-Chandrasekhar limit \citep{sc42} to be $M_{\rm c}\gtrsim (0.1$--$0.15) M_2$. Unless the star loses a large fraction of its mass {\it before\/} forming an He core, the minimum mass of the core is thus $\gtrsim 0.1\msun$. During the subsequent evolutionary stages, the core continuously gains mass and the stellar radius grows and remains $\gtrsim 1\rsun$, as can be seen from the dashed curves in Fig.~\ref{r_mc2}. Those minimum radii are significantly above the possible Roche-lobe radius in \source. The radius of the giant only shrinks when the star loses almost all the mass in its envelope (due to the H burning shell approaching the surface), see Fig.~\ref{r_mc2}. This quickly stop the accretion, given that accretion from the less massive star onto the more massive one leads to a widening of the orbit. The core becomes eventually a low-mass He white dwarf. 

The above considerations are also in agreement with the observational evidence that accreting LMXBs with giant donors have large periods (12\,d for GRO 1744--28, 34\,d for GRS 1915+105, and 6.5\,d for V404 Cyg), implying also relatively large Roche-lobe radii (see Eq.~\ref{r_k}). As the orbital period of \source\ is of only 6.3\,h, we conclude that a standard giant donor star cannot be hosted in this system.

Still, we cannot exclude the possibility that a giant has undergone a substantial mass loss before the formation of its He core. An argument in support of it is given by the existence of some LMXBs with very low-mass white dwarf donors (which were giant cores in the past), for example 4U 1820--303 with $M_2\simeq (0.06$--$0.08)\msun$ \citep{rappaport87}. In order to test this scenario, we have performed evolutionary calculations by exploiting again the Warsaw stellar evolutionary code, shown in Fig.~\ref{r_mc2}. To reproduce a system close to \source, we followed the evolution of a $1 \msun$ star, which was maintained at a constant mass until hydrogen was nearly exhausted in its centre. At this stage, the He core was not yet formed. \blue{Then, the mass removal from the surface started and it was continued until the donor star reached seven different values of the mass, 0.15, 0.16, 0.175, 0.2, 0.3, 0.5 and $0.9\msun$. The He core still did not form by that time. We applied two different (arbitrary) rates of the mass removal, $1.3\times 10^{-9}$, $1.3\times 10^{-8} \msun$/y. We have found that the structure of the remnants was identical in both cases. Our motivation was to obtain thermal-equilibrium remnants of a given mass. Since the structure of such remnants does not depend on their evolutionary history, the prescription for mass removal from the surface was unimportant.  The further evolution of the remnants was followed at a constant total mass. The H burning shell moves outwards, increasing the mass of the He core and decreasing that of the H-rich envelope. The radii of the partially stripped giants also increase with $M_{\rm c}$, but when the mass of their envelopes gets sufficiently low, they shrink, as shown in Fig.~\ref{r_mc2}.}

\blue{We find that the giant radius in the considered cases can be equal to the Roche-lobe radius only for $M_2\leq 0.2\msun$. Fig.~\ref{r_mc2} thus shows the Roche-lobe radii for $M_2=0.15$ and $0.2\msun$. The possible solutions for the case of \source\ are represented by the intersections between the evolutionary tracks and the Roche-lobe values on the left side of the plot. Solutions on the right side of the plot are not viable due to the fact that in this evolutionary stage the donor star radius shrinks rapidly within the Roche-lobe and the mass transfer toward the compact object cannot be sustained. The partially stripped-giant solutions presented here show that it is possible to form a very low-mass He core in the donors of short orbital period LMXBs, but only if the donor lost a substantial fraction of its mass before forming the core. Thus, if the donor in \source\ is a giant, we find the possible range of its mass of $M_2\simeq (0.16$--$0.20) \msun$. In this case, $R_2 \simeq (0.43$--$0.47)\rsun$ and, for an NS accretor, $i\simeq (79$--$81)\degr$. In the case of the BH accretor, $i\simeq 84$--$85\degr$ and 88--$89\degr$ for $M_1=5\msun$ and $\simeq\! 10\msun$, respectively. We have, however, neglected irradiation of the donor by the X-ray source, which may somewhat modify the above parameters.}

We note that \source\ shows both dips and nearly total eclipses \citep{bozzo16}. This is the case corresponding to an intermediate range of high inclinations, with values of $i$ lower than those of dip-only sources and higher than those for almost edge-on systems, where the disc obscures the central X-ray source \citep{frank87}. \citet{frank87}, based on considerations of the geometry of an accreting binary with an accretion disc corona, estimated the expected inclination range of binaries showing both dips and total eclipses as $\sim$(75--$80)\degr$. Unfortunately, even accepting this range at face value does not allow us to discriminate between the above solutions with either an MS star or a very low-mass giant except that we can exclude the case of a very low-mass giant accreting onto a BH (which has $i$ close to $90\degr$).

Partially stripped giants have no memory of their past evolution, and their structure is primarily determined by $M_{\rm c}$ and the total mass. In the standard case, with the total giant mass $\ga 0.3\msun$ or so, the stellar radius depends mostly on $M_{\rm c}$, with only a minor effect of the total mass \citep{paczynski70,rappaport95}, as we have already noted. However, we find here a new and interesting effect that the giant radii become strongly dependent on their total masses when they are low. This is illustrated in Fig.~\ref{r_mc2}, which shows that the dependence of the radii on the total mass becomes progressively stronger as the mass decreases. It is relatively weak between the masses of 0.9 and $0.3\msun$, but becomes highly significant for lower masses. This also shows that the approximate analytical formulae for $R_2(M_{\rm c})$ of \citet{webbink83} and \citet{king93} are no longer applicable in this regime.

We also note that our new evolutionary solutions for very low-mass giants could be applicable in the case of the LMXB X2127+119 in the globular cluster M15. In that object, with the orbital period of $P=17.1$\,h, \citet{vanzyl04a,vanzyl04b} found evidence for the presence of a very low-mass donor with $M_2\sim (0.1$--$0.15)\msun$. In this range, we still find no possible solutions, but at a slightly larger value of $M_2=0.175\msun$, the Roche-lobe radius of the donor would be $\simeq 0.87\rsun$, and our solution for this mass in Fig.~\ref{r_mc2} (the green curve) could be applicable for that source.

\blue{Our results are compatible with those of \citet{lin11}, who have performed extensive evolutionary calculations following evolution for binaries with a $1.4\msun$ neutron star down to the donor masses of $0.01\msun$. Their fig.\ 1 shows results fully consistent with ours. For $P\simeq 6.3$ h of \source, they obtain the allowed range of the donor mass of 0.15--$0.8\msun$ and 1.5--$2.5\msun$. The former range corresponds to that obtained by us, while the latter appears to correspond to very fast outflows with very short duration. In their fig.\ 2, two evolution tracks going through the period of \source\ are marked CV and UC (ultracompact). On the CV track, the donor star is only moderately evolved when the mass transfer starts, and it remains H-rich while losing mass by accretion. The  position of \source\ on this track corresponds to our solution with an MS donor. On the UC track, the mass transfer starts when the donor star is evolved, typically with H just having been depleted at the stellar centre and before formation of an He core. The position of \source\ on their UC track roughly corresponds to our solution with a partially stripped-giant donor. However, without the access to details of their calculations, we cannot use directly those results.}

Finally, we show that either brown or white dwarf donors in \source\ can be firmly ruled out. Old brown dwarfs are self-sustained by both the classical ionic-Coulomb pressure, dominating in at their low-mass range of $M\ga 0.012\msun$, and by the partial electron degeneracy, dominating at the high-mass range of $M\la 0.075\msun$. Their radii are close to $0.1\rsun$ in the entire mass range \citep{chabrier09}, which is far below any reasonable assumption for the size of the Roche-lobe in \source. We note that there has been a measurement of substantially higher radii in the double brown-dwarf binary 2MASS J05352184--0546085, namely $R/\rsun=0.669\pm 0.034$ at $M/\msun=0.054\pm 0.005$ and $R/\rsun=0.511\pm 0.026$ at $M/\msun=0.034\pm 0.003$ \citep{stassun06}. However, this is a recently formed system, with the age of $\sim\! 10^6$~y, orders of magnitude smaller than the typical age of LMXBs ($\sim\! 10^8$--$10^9$~y).

Concerning low-mass white dwarfs, their radii in the fully degenerate limit are given by $R/\rsun\simeq 0.0128(M/\msun)^{-1/3}(1+X)^{5/3}$ \citep{c39,db03}. A He or C/O white dwarf (with $X\simeq 0$) has thus the radius even smaller than that of a degenerate brown dwarf with a comparable mass (by the factor of $\sim\! (1+0.7)^{-5/3}\simeq 0.4$), and can be ruled out as well as the donor. An increase in the radius of an old white dwarf due to departures from degeneracy and a non-negligible H content can be only moderate, and would still result in a radius much smaller than the size of the Roche lobe in \source. On the other hand, young low-mass white dwarfs (i.e., shortly after their formation from cores of stripped giants) can be thermally bloated to significantly larger radii \citep{maxted11,maxted13,rappaport15}. However, the evolution of such objects is characterized by a fast shrinking radius (due to the white dwarf cooling), and thus they are unable to sustain a stable Roche-lobe overflow phase, as observed in \source.

\section{Conclusions}

We have constrained the binary parameters of \source\ using the recent measurements of the period and the eclipse duration of that system by \citet{bozzo16}. We have determined that the donor is most likely an MS star with the mass within $\sim$(0.5--$0.8)\msun$ and the radius of $\sim\! 0.7\rsun$. As found by \citet{bozzo16}, the properties of the source X-ray spectra suggest the presence of an NS accretor, even though a BH could not be firmly ruled out. In the NS case, we find the system inclination is $71\degr\lesssim i\lesssim 76\degr$.

We have also studied an alternative possibility that the donor is a very low-mass and small-radius partially stripped giant. This solution requires that an initially heavy progenitor of the donor (with $M_2\ga 1\msun$) lost a substantial amount of mass prior to the formation of its He core. From evolutionary calculations, we have found $M_2\sim (0.15$--$0.2)\msun$, $R_2\sim 0.4\rsun$, $i\sim 80\degr$. A similar solution could apply to X2127+119, where \citet{vanzyl04a,vanzyl04b} required the donor to have a very low mass. We have ruled out the cases of the donor in \source\ being either a standard (i.e., without major mass loss before the core formation) giant or a brown or white dwarf.

We have also obtained two general results. First, we have found that the radius of a giant depends sensitively on the giant mass at the masses $\la 0.3\msun$. This is unlike the standard, higher-mass case, where the radius depends mostly on the core mass, with only a secondary dependence on other parameters. We have also obtained an analytical solution, Eq.\ (\ref{incl}), of the standard eclipse equation, providing an explicit formula for the inclination as a function of the donor radius, the period, and the eclipse duration.

\begin{acknowledgements}
We thank Tomek Bulik, Phil Charles, Joanna Miko{\l}ajewska and Ashley Ruiter for valuable discussions, and the referee for valuable suggestions. This research has been supported in part by the Polish National Science Centre grants 2012/04/M/ST9/00780, 2013/10/M/ST9/00729 and 2015/18/A/ST9/00746.
\end{acknowledgements}

\bibliographystyle{aa}

\end{document}